# Understanding Repository Growth at the University of North Texas: A Case Study

Mark E. Phillips
University of North Texas
Denton, Texas 76205
mark.philips@unt.edu

Lauren Ko
University of North Texas
Denton, Texas 76205
lauren.ko@unt.edu

## ABSTRACT

Over the past decade the University of North Texas Libraries (UNTL) has developed a sizable digital library infrastructure for use in carrying out its core mission to the students, faculty, staff and associated communities of the university. This repository of content offers countless research possibilities for end users across the internet when it is discovered and used in research, scholarship, entertainment, and lifelong learning. The characteristics of the repository itself provide insight into the workings of a modern digital library infrastructure, how it was created, how often it is updated, or how often it is modified. In that vein, the authors created a dataset comprised of information extracted from the UNT Libraries' archival repository Coda and analyzed this dataset in order to demonstrate the value and insights that can be gained from sharing repository characteristics more broadly. This case study presents the findings from an analysis of this dataset.

## 1. INTRODUCTION

Understanding the characteristics and growth patterns in digital library infrastructure throughout the world is challenging. In many situations, digital library programs advertise the size of their repositories in relation to the number of items that are publically available or the number of pages that comprise the repository such as those given by HathiTrust [4]. These metrics are often aggregated in databases such as the OpenDOAR system [5], which allows users to obtain an understanding of the size of a repository. These metrics, however, are not always easy to obtain and organizations such as OpenDOAR [6] have had to do a non-trivial amount of work to provide accurate statistics for record counts. The number of digital objects available in a repository is only one of a number of metrics that are important in understanding the growth of a digital repository over time.

Additional metrics that may be of interest include the number of files that comprise a digital object, the size of the digital object, and the date that objects were added to the repository.

Another important factor to address is that often the public facing access side of a digital repository does not fully reflect the preservation infrastructure used to store the rich archival master files. In many organizations there are systems that maintain all files related to an object in one place, while other systems separate these two concepts and manage them differently.

The UNT Libraries operates a number of public facing digital library interfaces. These interfaces include The Portal to Texas History [8], the UNT Digital Library [9], and The Gateway to Oklahoma History [7] operated for the Oklahoma Historical Society. Each of these public interfaces makes available digital resources and associated metadata for consumption by users around the world. Digital objects are generally presented in Web deliverable formats such as JPEG, MP3, MP4, PDF, DOC, and so on. The interfaces have technology in place to present high-resolution images to the end user in a tiled zoom interface using either JPEG2000 or Zoomify Tiles. Each of the resources made available in these public interfaces is archived by the UNT Libraries in its digital archival repository called Coda. There are digital objects in the Coda repository that do not have direct Web accessible components at this time, including large research datasets and digital objects that are stored in a "dark" mode for the time being.

At the time of writing this paper, there are 688,427 total digital objects in the public facing digital collections system with 642,155 currently accessible by users. In the digital archival repository Coda, there are 689,628 digital objects occupying 150 TB of storage and accounting for 90,830,322 files. The analysis in this paper is based on the digital objects held in the Coda repository.

## 2. RESEARCH QUESTIONS

The following research questions drove this analysis.

- What are the growth trends in digital content acquisition for the UNT Libraries' Digital Collections?

- What are the average characteristics in relation to size, scale, and frequency of ingest of digital objects?

- How has the versioning strategy chosen by the UNT Libraries been utilized in the digital collections?

## 3. BUILDING THE DATASET

The UNT Libraries stores its primary digital objects in a purpose built digital archival system developed by the UNT Libraries called Coda. This system is responsible for ingest, monitoring, and dissemination of the rich digital master files that make up the UNT Libraries Digital Collections. The Coda system has been in full operation at UNT since September of 2009 when the first archival objects were ingested. The primary object model for the Coda system is a BagIt bag. Coda makes no assumption of the contents of the bag deposited into the system and utilizes the features of the BagIt specification [2] to audit, verify, and ingest each bag into the system. The UNT Libraries has made the decision that each intellectual unit added to the UNT Libraries Digital Collections will be contained within its own bag, whether it be a photograph, newspaper issue, book, audio or video recording, ETD, website or a dataset. This allows for a logical mapping of a physical or born digital object to the archival version of the object in the Coda repository.

The dataset used in this study contains information about each digital object stored in the Coda system. Table 1 shows the data elements that are available for each record instance in the dataset.

Table 1. Data elements available per record

| Field | Sample Value | Description |
| --- | --- | --- |
| Coda Identifier | ark:/67531/coda4w | Unique identifier created by the Coda system upon ingest |
| File Count | 43 | Number of files in the /data/ directory of the BagIt bag |
| Bytes | 73751840 | Number of bytes in the /data/ directory of the BagIt bag |
| Ingest Date | 2009-09-24 | Date in which the bag was added to Coda |
| External Identifier | ark:/67531/metapth47242 | External identifier for the bag used in the public access systems |

A total of 682,745 records were extracted from the Coda system with Ingest Dates from September 24, 2009 through March 31, 2014. The authors added each to a tab delimited file and sorted the records by Ingest Date. This tab-delimited file was used by the authors for the analysis in this paper.

## 4. ANALYSING THE DATA

The authors utilized both common command-line tools to parse and calculate statistics as well as Python scripts to aggregate and group subsets of records for analysis. The findings for this paper represent a case study on what content was added to the UNT Libraries' Coda repository from September 2009 until the end of March 2014. The analysis is split into several sections, first, an analysis by year of the dataset, which explores the number of objects, added, growth of disk space, and accumulation of files. Second, the authors look at the characteristics of the objects themselves, including statistics based on the number of files per digital object and the amount of disk space used per object. Finally, a review of the methodology of versioning employed by the UNT Libraries is discussed with the authors identifying the overhead storage resources consumed by this versioning strategy.

## 5. AGGREGATE DATA BY YEAR

The number of digital objects added to the Coda repository each year is presented in Table 2. The average number of digital objects added to the Coda repository each year is 113,790, it should be noted that 2009 and 2014 represent only four and three months of data respectively.

Table 2. Number of objects added per year

| Year | Objects |
| --- | --- |
| 2009 | 21,992 |
| 2010 | 134,340 |
| 2011 | 111,990 |
| 2012 | 154,895 |
| 2013 | 210,089 |
| 2014 | 49,439 |

The average number of files added to the repository each year is 14,944,060 files, again with 2009 and 2014 being low in the dataset because of the truncated periods of data. The annual totals are shown in Table 3 below.

Table 3. Number of files added per year

| Year | Files |
| --- | --- |
| 2009 | 776,429 |
| 2010 | 29,013,866 |
| 2011 | 19,500,976 |
| 2012 | 19,638,101 |
| 2013 | 15,859,711 |
| 2014 | 4,875,274 |

The dataset shows amount of storage utilized by the Coda repository and how it has grown over the past six years. The average amount of data added to the repository is just over 25 TB of data per year. The breakdown per year is presented in Table 4.

Table 4. Number of GB added per year

| Year | Gigabytes |
|---|---|
| 2009 | 2,522 |
| 2010 | 44,708 |
| 2011 | 18,535 |
| 2012 | 25,624 |
| 2013 | 34,067 |
| 2014 | 26,868 |

Because the dataset used in the analysis contains a truncated year at the beginning of the dataset and at the end, the authors felt that an analysis offering a monthly view of the growth in the Coda repository would provide additional insight into the growth of this system. Table 5 presents this data with monthly averages of the number of objects, files, and associated size in GB of data added to the repository. This analysis helps to minimize the effects of the truncated data from 2009 and 2014.

Table 5. Monthly averages of the number of objects, files per object, and object size.

| Year | Average Monthly Objects | Average Monthly Files | Average Monthly Size |
|---|---|---|---|
| 2009 | 5,498 | 194,107 | 631 GB |
| 2010 | 11,195 | 2,417,822 | 3,726 GB |
| 2011 | 9,332 | 1,625,081 | 1,545 GB |
| 2012 | 12,907 | 1,636,508 | 2,135 GB |
| 2013 | 17,507 | 1,321,642 | 2,839 GB |
| 2014 | 16,479 | 1,625,091 | 8,956 GB |
| All Years | 12,413 | 1,630,261 | 2,770 GB |

## 6. OBJECT CHARACTERISTICS

The dataset offers the authors two ways of working with the size of the digital objects in the repository, both by the number of files and the size of the digital object.

Because the dataset analyzed contains the number of files that make up each digital object, the authors used these values to identify the average number of files that comprise a digital object in the UNTL collection.

Table 6. Files per object by year

| Year | N | Min | Max | Mean | Stddev | Mode |
|---|---|---|---|---|---|---|
| 2009 | 21,992 | 4 | 6,850 | 35 | 227.6 | 4 |
| 2010 | 134,340 | 3 | 31,741 | 216 | 885.8 | 4 |
| 2011 | 111,990 | 3 | 23,605 | 174 | 851.2 | 4 |
| 2012 | 154,895 | 3 | 51,908 | 127 | 646.0 | 98 |
| 2013 | 210,089 | 3 | 24,418 | 75 | 317.8 | 6 |
| 2014 | 49,439 | 3 | 12027 | 99 | 368.8 | 6 |
| All Years | 682,745 | 3 | 51908 | 131 | 643.1 | 6 |

The authors were interested in understanding the sizes of digital objects added to the Coda system over time. Digital object sizes were grouped into ranges from under one Megabyte to digital objects that are larger than one Gigabyte. Table 7 presents this data at both the yearly aggregation as well as all of the years combined.

Table 7. Object by size, counts by year

| Year | Objects | < 1 MB | 1 MB - 100 MB | 101 MB- 1 GB | > 1 GB |
|---|---|---|---|---|---|
| 2009 | 21,992 | 5 | 16,797 | 5,060 | 130 |
| 2010 | 134,340 | 3,156 | 71,517 | 57,169 | 2,498 |
| 2011 | 111,990 | 22,813 | 52,815 | 35,505 | 857 |
| 2012 | 154,895 | 13,829 | 62,339 | 77,341 | 1,386 |
| 2013 | 210,089 | 2,144 | 119,470 | 85,219 | 3,256 |
| 2014 | 49,439 | 254 | 39,413 | 8,647 | 1,125 |
| All Years | 682,745 | 42,201 | 362,351 | 268,941 | 9,252 |

Another view of this data is presented in Table 8 where the percentage of the total number of objects contained in the size category is shown instead of the absolute count. This helps the user see that for the whole repository collection, 6.2 percent of the digital objects are less than 1 Megabyte in size, and a very small number of objects, 1.4 percent, are greater than 1 Gigabyte in size. The vast majority (92.5 percent) of the digital objects in the repository range in size from 1 Megabyte to 1 Gigabyte in size.

Table 8. Object size percentages by year

| Year | Objects | < 1 MB | 1 MB-100 MB | 101 MB-1 GB | > 1 GB |
|---|---|---|---|---|---|
| 2009 | 21,992 | 0.0% | 76.4% | 23.0% | 0.6% |
| 2010 | 134,340 | 2.3% | 53.2% | 42.6% | 1.9% |
| 2011 | 111,990 | 20.4% | 47.2% | 31.7% | 0.8% |
| 2012 | 154,895 | 8.9% | 40.2% | 49.9% | 0.9% |
| 2013 | 210,089 | 1.0% | 56.9% | 40.6% | 1.5% |
| 2014 | 49,439 | 0.5% | 79.7% | 17.5% | 2.3% |
| All Years | 682,745 | 6.2% | 53.1% | 39.4% | 1.4% |

## 7. ANALYSIS OF VERSIONS

In 2009, the UNT Libraries made the decision to manage versions of digital objects in the following way. As a change is made to the digital object that would require the files to change, the object is versioned by adding the object into the Coda repository as a new object. The Coda repository will assign the object a new internal Coda Identifier while maintaining External Identifiers submitted with the digital object. By analyzing the number of digital objects that are in the Coda repository that have multiple versions, we can understand how much the digital objects change and to what effect these changes have on underlying resources such as storage because each version of an object is stored multiple times.

Of the 682,745 digital objects represented in the dataset, there are 681,208 unique digital objects. These digital objects are identified by the External Identifier submitted with the digital object upon ingest. There are 1,527 digital objects (0.2 %) that contain multiple versions in the Coda repository. Of these 1,527 digital objects, 1,517 have two versions and 10 have three versions of the digital object.

The overhead incurred because of the redundant storage of these objects is 346.5 Gigabytes out of the 152,324 Gigabytes of storage consumed by the Coda repository, this overhead accounting for 0.2 % of the total storage utilized by the system.

## 8. DISCUSSION

The ability to track the growth of a digital repository over time is helpful in the prediction of growth and provisioning of storage resources. While many institutions are investigating storage options that allow for on-demand acquisition of storage resources from cloud service providers, the UNT Libraries has found that for the primary and secondary copies of their digital repository master files, local operation of storage infrastructure is more cost effective than cloud based solutions. The ability to take historical data and create trends for the growth can provide a baseline storage growth pattern to inform storage infrastructure purchases.

Another aspect of this analysis that the authors found interesting was in relation to versioning of digital objects and the overhead incurred within the repository because of the chosen versioning strategy. There has been much investigation into various digital object versioning strategies, most notably the work by Richard Anderson at Stanford Libraries [1]. The team at the UNT Libraries made the decision to adopt the most basic versioning strategy where when an object changes, a complete copy of the new version of the resource is added to the repository. This results in overhead in the storage costs because of the duplication, but as the authors noted above, this overhead amounts for only 346.5 Gigabytes of storage. A further description of the notion of "change" in the UNT context is needed for a thorough understanding of the implementation. The UNT Libraries manages the editing and versioning of descriptive metadata records associated with a resource in a separate system. These changes happen across the digital library infrastructure several hundred times a day and are stored and versioned in a metadata editing system. Digital objects in the Coda repository are linked to these metadata records using the external Archival Resource Key (ARK) [3] identifier included with a digital object upon ingest into the Coda repository. Only changes to the files within a digital object are introduced into the Coda repository. Examples of these changes include reordering of files, insertion of a file, removal of a file, or modification of an existing file.

The authors thought it would be interesting to show how compactly portions of the UNT Libraries' Digital Collections could be stored if one were to fill different sized containers, or "buckets" as shown in Table 9. The authors found it would be possible to fit almost 50 percent of the repository onto a single six Terabyte hard drive. This demonstrates that a large number of the digital objects in the repository are small in size, while a small number of objects are quite large in size. Put another way, the top 20 percent of objects based on size (136,549 objects of 682,745) account for 79 percent of the storage required for the repository (119,630 GB of 152,324 GB).

Table 9. Number of objects that will fit within different size "buckets"

|  | 1 MB | 1 GB | 1 TB | 6 TB* |
|---|---|---|---|---|
| **Count of objects** | 75 | 13,483 | 175,633 | 340,010 |
| **Percentage of repository** | 0.01 % | 1.97% | 25.7% | 49.8% |

*Largest consumer hard drive available at the time of writing

## 9. FURTHER WORK

It would be beneficial to the community if other digital repositories would create and make available datasets similar to the dataset used in this paper so that comparative analysis could be performed. Without sufficient data from other institutions, it is impossible to know if the data and analysis presented by this case study of the UNT Libraries Digital Collections is unique or similar to that of other institutions. Additionally, the collection of datasets like this would make it possible to analyze this data over time in longitudinal studies which may uncover new trends that can help with the planning of repository services in the future.

## 10. DATASET

The authors have placed the dataset used in this paper into the public domain with a CC0 1.0 Universal (CC0 1.0) Public Domain Dedication. The dataset is made available via the UNT Digital Library.

If you are interested in utilizing this dataset for similar work, please cite it as follows:

*Phillips, Mark and Ko, Lauren. (2014) Coda Archival Digital Repository Dataset. UNT Digital Library. http://digital.library.unt.edu/ark:/67531/metadc282640.*